\documentclass[11pt]{article}
\pdfoutput=1

\usepackage{amsmath,amssymb,bm,color}
\usepackage{appendix}
\usepackage{graphicx}
\usepackage{xcolor}
\usepackage{cancel}
\usepackage{jcappub}
\usepackage[T1]{fontenc} 
\usepackage{slashed}
\usepackage{soul}

\usepackage{epsfig}
\usepackage{array}
\usepackage{epstopdf}
\usepackage{bm}
\usepackage{tikz}
\usepackage{cancel}
\usepackage{xcolor}
\usepackage{float}
\usepackage{graphicx}
\usepackage{mathrsfs}
\usepackage{slashed}
\usepackage{textcomp}
\usepackage{times}
\usepackage{subcaption}
\usepackage{url}
\usepackage{cancel}
\usepackage{textcomp}
\usepackage{amsmath, amssymb, amsfonts, latexsym, epsfig}
\usepackage{mathrsfs}
\usepackage{hyperref}

\newcommand{\fGW}{f_{\rm GW}}

\def\beq{\begin{equation}}
\def\eeq{\end{equation}}
\def\bea{\begin{eqnarray}}
\def\eea{\end{eqnarray}}

\def\sun{\odot}

\def\lsim{\mathrel{\rlap{\lower4pt\hbox{\hskip1pt$\sim$}}
     \raise1pt\hbox{$<$}}}         
\def\gsim{\mathrel{\rlap{\lower4pt\hbox{\hskip1pt$\sim$}}
     \raise1pt\hbox{$>$}}}         

\newcommand{\be}{\begin{equation}}
\newcommand{\ee}{\end{equation}}

\title{\boldmath Exploring Dark Forces with Multimessenger Studies of Extreme Mass Ratio Inspirals}

\author[a]{Badal Bhalla,}

\author[a]{Kuver Sinha,}

\author[a]{and Tao Xu}

\affiliation[a]{Homer L. Dodge Department of Physics and Astronomy, University of Oklahoma, Norman, OK 73019, USA}

\emailAdd{badalbhalla@ou.edu}

\emailAdd{kuver.sinha@ou.edu}

\emailAdd{tao.xu@ou.edu}

\abstract{The exploration of dark sector interactions via gravitational waves (GWs) from binary inspirals has been a subject of recent interest. We study dark forces using extreme mass ratio inspirals (EMRIs), pointing out two issues of interest. Firstly, the innermost stable circular orbit (ISCO) of the EMRI, which sets the characteristic length scale of the system and hence the dark force range to which it exhibits enhanced sensitivity, probes force mediator masses that complement those studied with supermassive black hole (SMBH) or neutron star binaries. The LISA mission (the proposed $\mu$Ares detector) will probe mediators with masses $m_V \sim 10^{-16}~{\rm eV}$ ($m_V \sim 10^{-18}~{\rm eV}$), corresponding to ISCOs of $10^6 M_\odot$ ($10^8 M_\odot$) central SMBHs. Secondly, while the sensitivity to dark couplings is typically limited by the uncertainty in the binary component masses, independent mass measurements of the central SMBH through reverberation mapping campaigns or the motion of dynamical tracers enable one to break this degeneracy. Our results therefore highlight the necessity for coordinated studies, loosely referred to as ``multimessenger", between future $\mu{\rm Hz}-{\rm mHz}$ GW observatories and ongoing and forthcoming SMBH mass measurement campaigns, including OzDES-RM, SDSS-RM, and SDSS-V Black Hole Mapper.
}

\begin{document}
\hfill{\small CETUP-2023-013}

\maketitle
\flushbottom

\section{Introduction}

A wide range of physics scenarios beyond the Standard Model (SM) and General Relativity (GR) have as their template the following potential generated by an object of mass $M$:
\bea \label{introyovan}
U \, = \, - \frac{G M}{r}\left[1+ \tilde{\alpha}' e^{-\frac{r}{\lambda}}\right],
\eea
where the extra contribution to the Newtonian potential, sometimes dubbed a ``fifth force" or ``dark force" depending on the context, is controlled by an effective coupling strength parameter $\tilde{\alpha}'$ and a characteristic length scale $\lambda$. The Yukawa contribution can also be interpreted as the static limit of an interaction mediated by bosons of mass 
\be
m_V \, \sim \, \frac{1}{\lambda} \,\,.
\ee
This widely used framework has historically served to parametrize deviations from GR and the inverse square law \cite{deRham:2016nuf, Adelberger:2003zx, Clifton:2011jh}, the effects of light string moduli \cite{Brzeminski:2022sde, Damour:2010rm, Acharya:2018deu, Kane:2015jia} and, in more recent times, as a template to study light mediators in  dark sector models \cite{Croon:2017zcu, Dror:2021wrl, Kopp:2018jom, Alexander:2018qzg, Poddar:2023pfj, Gresham:2022biw}. Interpreted purely in terms of deviations from GR (i.e., without reference to dark sector physics), this formalism describes, for example, models of large extra dimensions (for small values of $\lambda \ll 10^{-3}$m) \cite{Kehagias:1999my, Kaplan:2000hh} as well as braneworld~\cite{Callin:2004py, Fichet:2022ixi, Fichet:2022xol}, and the linearized regime of massive gravity theories (for large values of $\lambda$) \cite{deRham:2016nuf}. In the study of dark sectors, on the other hand, $\tilde{\alpha}'$ takes on the role of the dark sector coupling and $m_V$ the role of the dark mediator mass, while the Yukawa interaction is assumed to be sourced by the accumulation of dark charge on the central and secondary objects. Indeed, a host of interaction potentials (not just the Yukawa potential) are possible, depending on the nature of the mediator being exchanged (we refer to \cite{Brax:2017xho, Banks:2020gpu} for a unified treatment of such exchange interactions, and \cite{Costantino:2019ixl, Xu:2021daf} for recent work along these lines). We will mainly restrict our attention to long-ranged Yukawa potentials in this work. 

The phenomenological framework of Eq.~\eqref{introyovan} encapsulates a wide variety of theories. For the purposes of this work, it is useful to distinguish between $(i)$ theories of modified gravity or fifth forces, which would apply to all bodies; and $(ii)$ theories of dark sector physics, where Eq.~\eqref{introyovan} only applies to bodies with dark charge. While our main interest is in the second category, we briefly summarize the various systems that have been used to constrain fifth forces in general. The parametrization in Eq.~\eqref{introyovan} dictates the use of diverse experimental settings, where the characteristic scale of the probe should be $\lambda_{\rm probe} \sim \lambda$. The pursuit of probing $\lambda$ through measurements at laboratory scales has a long history \cite{Talmadge:1988qz}. In the large $\lambda$ (light mediator) regime,  competitive bounds arise from  planetary motion and the motion of stars around Sgr A$^*$ (we refer to \cite{Talmadge:1988qz} and \cite{Johannsen:2015mdd}  for  comprehensive treatments of these two frontiers). The most stringent Solar System constraints on the $(\tilde{\alpha}', \lambda)$ plane come from modifications of Kepler's Third Law. On the other hand, broadly three avenues have been pursued with respect to constraints from Sgr A$^\star$ \cite{Hees:2017aal}: $(i)$  Orbital precession data  by  near infrared monitoring of stars  that are sufficiently close to Sgr A$^\star$;  $(ii)$  high-precision timing of pulsars by existing and future radio telescopes; and $(iii)$ space- and time-resolved probes of the accretion flow by the Event Horizon Telescope. The advent of gravitational wave (GW) astronomy has opened up new opportunities in this frontier (we refer to  \cite{Yunes:2013dva} for a recent review)\footnote{It should be noted that many of the  constraints on the $(\tilde{\alpha}', m_V)$ plane coming from GW data are obtained under the specific assumption that Eq.~\eqref{introyovan} depicts theories of massive gravitons (the constraints come from  corrections to the propagation of GWs from the source binary to Earth, due to the non-zero graviton mass). For example, very strong constraints on $m_V$ are obtained by a comparison of the times of arrival of GW and electromagnetic signals, or pure GW signals coming from sky-averaged, quasi-circular inspirals of compact objects of various masses (we refer to \cite{Will:1997bb} for an early landmark paper in this area, and Table 2 of  \cite{Yunes:2013dva} for a summary of constraints). The  use of GW  propagation to constrain massive gravitons is \textit{not} the focus of our work.}.

Compact object mergers in particular present unique opportunities: the population distributions and the GW waveforms emanating from these systems are sensitive to modifications of the type parametrized in Eq.~\eqref{introyovan}. Different classes of compact binary systems enable different scales $\lambda$ to be probed, depending on the characteristic frequency of the GW emission. The dark sector interpretation of Eq.~\eqref{introyovan}, which is our main interest, works 
under the assumption that both the interacting objects accumulate dark charge. In the case of inspiraling neutron stars, such accumulation could result from dark matter capture; in the case of solar or super massive black holes, such accumulation could result from spike formation, or from the fact that such black holes are charged under millicharged dark fermions \cite{Cardoso:2016olt}. Here, too, the recent literature is substantial, and we refer to \cite{Croon:2017zcu, Dror:2021wrl, Kopp:2018jom, Alexander:2018qzg, Poddar:2023pfj} for some representative studies.

The purpose of this paper is to explore the possibility of probing dark forces with extreme mass ratio inspiral (EMRI) systems consisting of a central object -- the super-massive black hole (SMBH) -- and a stellar mass  black hole in a decaying orbit around the central object \cite{PhysRevD.95.103012, Berry:2019wgg}. These systems produce millions of cycles in the GW sensitivity band of LISA, enabling probes of the background geometry, as well as the detection of small changes to the radiation-reaction force. The inspiral process is driven mainly by the emission of gravitational radiation, thus providing an excellent laboratory in which to probe any altered emission due to dark forces. EMRI systems have been extensively used as laboratories of  GR and modified gravity: we refer to \cite{Barsanti:2022vvl, Maselli:2021men} for papers relevant to our study. Their use as probes of dark forces and dark sectors is somewhat more limited and recent: we refer to \cite{Barsanti:2022vvl, Maselli:2021men} for scalar-mediated dark force studies, and \cite{Dror:2021wrl} for somewhat related dark sector studies of SMBH binaries.

There are two main reasons why EMRIs should be studied in this context. The first is that the innermost stable circular orbit (ISCO) of the EMRI, which determines the characteristic length scale of the system and thus the characteristic  wavelength of the mediator, probes a dark force range that is complementary to that probed by SMBH or neutron star binaries. Moreover, since the ISCO scales linearly with the mass of the central SMBH, different EMRI systems are sensitive to different dark force ranges corresponding to $\lambda \, \sim \, R_{\rm ISCO} \, \propto \, M_2$, where $M_2$ is the mass of the SMBH. In particular, we show our results for LISA~\cite{LISA:2017pwj, Colpi:2024xhw} (SMBH masses $\sim 10^6 \, M_\sun$ with corresponding $\lambda \sim \mathcal{O}(10^{10})$ m and $m_V \sim \mathcal{O}(10^{-16})$ eV) and for the proposed $\mu$Ares detector~\cite{Sesana:2019vho} (SMBH masses $\sim 10^8 \, M_\sun$ with corresponding $\lambda \sim \mathcal{O}(10^{12})$ m and $m_V \sim \mathcal{O}(10^{-18})$ eV).

The second reason EMRIs are interesting in this context is that mass measurements of the central SMBH, performed by methods that are independent of the primary EMRI system (a combination that we somewhat loosely call ``multimessenger") can enable us to probe small values of the dark sector coupling. The main challenge in probing small couplings $\tilde{\alpha}'$ is that the correction in Eq.~\eqref{introyovan} is parametrically limited by the uncertainty in the binary component masses\footnote{A possible remedy that has been pursued by some authors  \cite{Seymour:2020yle} is to ascertain the masses of the binary components with post-Keplerian  parameters such as advance rate of periastron and Shapiro time delay. However, this is predicated on the assumption that the dark sector contribution to the orbital decay proceeds through a dominant dipole term (while the  contribution to  other post-Keplerian parameters comes from corrections to the potential in Eq.~\eqref{introyovan}). Rephrased in terms of post-Newtonian counting, the dipole radiation correction to the orbital decay  enters at -1PN order compared to GR, while the  corrections to other post-Keplerian  parameters enter at 0PN.}. Independent mass measurements  can break the degeneracy between gravity and the dark force, if the dark force is assumed to be  subdominant or ``switched off" between the SMBH and ordinary matter, while being operational between the SMBH and the black hole comprising the EMRI system. Stated in another way, the central SMBH is envisioned to have ordinary matter in its vicinity, which allows for the  determination of the mass of the SMBH, without the confounding effect of the dark force. This knowledge of the mass of the SMBH can then be used to constrain the dark force using the EMRI between the SMBH and stellar black hole. Although we will be somewhat agnostic to the underlying particle physics model that achieves this, we will indicate the most natural  setting where our scenario may be applicable: a dark sector fermion $\chi$ with very feeble millicharge (near the weak gravity conjecture limit) interacting via an Abelian gauge boson $V_\mu$. 

Measuring the mass of SMBHs is a topic of intense research in astronomy. In quiescent galaxies, the methods typically rely on the motion of dynamical tracers -- stars or gas -- combined  with sophisticated dynamical models. On the other hand, for active galaxies with prominent broad emission lines, the most promising method is reverberation mapping \cite{1982ApJ...255..419B,  1993PASP..105..247P, 2019MNRAS.487.3650H}. While measuring the motion of dynamical tracers requires high resolution and is applicable for nearer galaxies, reverberation mapping can be applied to galaxies at greater  distances. We will also consider a third method that is applicable to nearby galaxies: GW signals from extremely large mass-ratio inspirals (XMRIs) formed by the central SMBH and a brown dwarf. We note that there are other emerging proposals for measuring the mass of SMBHs that we do not consider, such as  X-ray flux variability of active galactic nuclei \cite{Akylas:2022dbv}. Our work motivates cross-pollination between the exciting field of SMBH mass determination and the future field of EMRI studies with future GW detectors, all within the context of dark sector physics.

The rest of the paper is organized as follows. In Section \ref{sec:multimessenger}, we discuss multimessenger measurements of the SMBH mass in different astrophysical systems. In Section \ref{sec:EMRIGW}, we show the dark force effect in the GW emission from EMRIs and the resulting sensitivity to the dark force with future LISA and $\mu$Ares observations. We conclude in Section \ref{sec:conclusion}.

\section{EMRIs, Multimessenger Studies, and Ultralight Mediators }
\label{sec:multimessenger}

For ultralight mediators, the prospects of distinguishing the dark force from pure gravity is challenging. Compared to equal mass binaries, however, EMRIs offer certain advantages in addressing this important problem. The key point is the determination of the mass of the SMBH by a method for which the dark force can be switched off, followed by the determination of the mass of the mediator by an EMRI between the SMBH and  a stellar black hole.   

\subsection{Multimessenger Studies with XMRI-EMRI Combinations}

The combination of an extremely large mass-ratio inspiral (XMRI) and an EMRI with the same central SMBH but two different smaller objects (a brown dwarf for the XMRI \cite{Amaro-Seoane:2020zbo, Vazquez-Aceves:2022dgi, Amaro-Seoane:2019umn} and a black hole for the EMRI) constitutes our  first example of a multimessenger signal. XMRIs have been recently explored as interesting sources for LISA or TianQin. With mass ratios as small as $\sim 10^{-8}$, a brown dwarf would complete $\sim 10^8$ cycles before the merger, spending time on band for millions of years. These are expected to be abundant and loud  systems: $10^4$ ($10^3$) years before merger, the SNR is expected to be $\mathcal{O}(10^3)$ ($\mathcal{O}(10^4)$) \cite{Amaro-Seoane:2019umn}. 
It should be noted that the high SNR of XMRI systems implies that they can be detected in nearby galaxies as well (\cite{Amaro-Seoane:2019umn} calculates that XMRI systems with SNR $= 10$ can be detected out to 50 Mpc  200 yrs before plunge). It is instructive to consider  the case of Sgr A$^\star$. XMRIs between Sgr A$^\star$ and brown dwarfs can constrain the mass of Sgr A$^\star$ at the level of $\mathcal{O}(10^{-2}) M_\sun$ \cite{Vazquez-Aceves:2022dgi} \footnote{We will assume that the redshifted SMBH mass and the physical SMBH mass have the same level of uncertainty in this case. This is possible if the redshift is accurately determined, as expected for concurrent XMRI-EMRI events.}. For the event rate of eccentric XMRIs, \cite{Vazquez-Aceves:2022dgi} obtain an average of $N = 8^{+9}_{-3}$ for the case of low spin of Sgr A$^\star$ ($a=0.1$) and $N = 12^{+6}_{-4}$ for the case of high spin of Sgr A$^\star$ ($a=0.9$). The numbers quoted above signify sources that emit GWs in the LISA and TianQin detection band with SNR $\sim 30$. The brown dwarfs are assumed to have mass $0.05 M_\sun$ in prograde orbits and have an orbital inclination of $0.1$ rad. The initial eccentricity of the XMRI system is taken to be between $0.99$ and $\geq 0.999$ with semimajor axis $4 \times 10^{-4}$ pc $\leq \, a \, \leq 8 \times 10^{-3}$ pc. Similar numbers are reported by \cite{Amaro-Seoane:2019umn}. 

We now turn to a discussion of EMRI formation. Generally, the  rate of EMRI formation in a galaxy is determined by two timescales. The first (the two-body relaxation timescale $T_{\rm rlx}$) is the time it takes to produce high eccentricity EMRI candidates from the dense cluster of stars and stellar remnants in the nuclear cluster surrounding the SMBH. The timescale $T_{\rm rlx}$ is determined by the number density of compact objects in the nuclear cluster, their velocity dispersion, steepness profile, and half mass radius \cite{1987gady.book.....B}. The second timescale pertains to the time it takes for the EMRI candidate to decouple from the rest of the nuclear cluster, and is determined by the rate of GW emission. Typical estimates for the relaxation time are in the range of $\mathcal{O}(0.1-1)$ Gyr. Combining the estimates of these timescales, the benchmark figure for EMRI event rates from studies in the literature is $\sim 10^{-6}$ per galaxy per year. This translates to tens to hundreds of EMRI events per year at LISA, which probes the Universe out to redshift $z \sim 1$ \cite{2009astro2010S.209M}. It should be noted that the event rate for EMRIs is a subject of ongoing research, and the benchmark rate quoted above can be boosted by other mechanisms. While the $\sim 10^{-6}$ per galaxy per year is based on the ``loss cone" scenario where a candidate is scattered by interactions with other bodies in the nucleus into a highly eccentric orbit around the SMBH, the presence of an accretion disk can  significantly enhance this process. For example, \cite{Pan:2021ksp} find that accretion disk-assisted EMRIs form $\sim \mathcal{O}(10^1 - 10^3)$ times faster than the ``loss cone” mechanism.

The possibility of a concurrent observation of an XMRI-EMRI combination at Sgr A$^\star$ is pessimistic: while the expected event rate of an XMRI is relatively high, the event rate of an EMRI is  more suppressed, and the joint probability is further suppressed.  Nevertheless, given that XMRI signals may be detected up to 50 Mpc, and EMRIs even further, the possibility of a coincident XMRI-EMRI signal may be significantly boosted if one considers other galaxies. Since such a concurrent multimessenger observation represents one of the best case scenarios for constraining ultralight  mediators, we depict  results corresponding to this benchmark scenario in our plots in Section \ref{sec:EMRIGW}.

\subsection{Multimessenger Studies with Stellar Kinematics - EMRI Combinations}

We now turn to another possibility: multimessenger combinations from stellar kinematics  around SMBHs on the one hand, and EMRIs of stellar black holes  on the other. While the precision of mass measurements using stellar orbits is far lower than that using the XMRI-EMRI combination, the event rate of such a possibility is more optimistic.

For the case of Sgr A$^\star$, the mass measurement by the GRAVITY Collaboration \cite{GRAVITY:2023avo, 2018A&A...618L..10G, GRAVITY:2018ofz} relies on stellar orbits at several thousands of gravitational radii and yields a value of $M_2 = (4.297 \pm 0.012) \times 10^6\, M_\sun$. On the other hand, \cite{2019A&A...625A..62T} provide six SMBH mass measurements at $3\sigma$ confidence level from SINFONI observations of nearby fast rotating early-type galaxies: NGC 584 ($M_2 = (1.3 \pm 0.5) \times 10^8\, M_\sun$), NGC 2784  ($M_2 = (1.0 \pm 0.6) \times 10^8\, M_\sun$), NGC 3640 ($M_2 = (7.7 \pm 5) \times 10^7\, M_\sun$), NGC 4570 ($M_2 = (6.8 \pm 2.0) \times 10^7\, M_\sun$), NGC 4281 ($M_2 = (5.4 \pm 0.8) \times 10^8\, M_\sun$) and NGC 7049 ($M_2 = (3.2 \pm 0.8) \times 10^8\, M_\sun$). Other SMBH mass measurements are also available: the MASSIVE Survey \cite{Ma_2014} is a volume-limited, multi-wavelength, integral-field spectroscopic and photometric survey of $\mathcal{O}(100)$ of the most massive early-type galaxies within 108 Mpc. A sample of 72 SMBHs and their host galaxies based on kinematic data and modeling efforts is provided in~\cite{McConnell_2013}. We note that the general level of the uncertainty in the mass measurement using stellar kinematics is $\sim \mathcal{O}(10-30)\%$. Thus, the observation of an EMRI between the SMBH and a stellar black hole in any of these galaxies would result in  multimessenger combinations of stellar kinematics-EMRI.

\subsection{Multimessenger Studies of Spectroscopic Reverberation Mapping and  EMRIs}

The third multimessenger combination scenario that we will consider is the measurement of the SMBH mass with spectroscopic reverberation mapping. In this method, the time delay in changes in the continuum emission arising from the accretion disk near the SMBH and the response of these changes on the emission from the (farther away) broad line region are measured. This time delay effectively measures the distance between the accretion disk and the broad line region (more precisely, the responsivity-weighted average  radius of the broad line region). Finally, from this information, the mass of the SMBH can be extracted following \cite{Peterson_2004}: 
\begin{equation}
M_2 \, = \, \frac{f c \tau_{\rm cent} \sigma^2_{\rm line}}{G} \,\,.
\end{equation}
Here, $f \sim \mathcal{O}(1)$ is a parameter that is specific to the broad line region and depends on its structure, orientation, and  kinematics. The parameter $\tau_{\rm cent}$ is the measured time delay (as measured from the centroid of the cross-correlation function around the emission peak), while $\sigma_{\rm line}$ denotes the root-mean square width of emission in the variable part of the spectrum. A major source of uncertainty is the value of $f$. Mass measurements of  SMBHs using this method can be obtained from the AGN Black Hole Mass Database \cite{2015PASP..127...67B}. Most of the measured SMBH masses lie at redshifts $z \, < \, 0.2$, although there are a few at $z \, \sim \, \mathcal{O}(2-3)$. The general level of uncertainty in the mass measurement is $\sim \mathcal{O}(10\%)$.

In this context, we take particular note of the OzDES and SDSS reverberation Mapping Programs \cite{2023MNRAS.520.2009M, 2019MNRAS.487.3650H, Shen:2023pht}, which have pioneered large-scale  reverberation mapping campaigns  \cite{2023MNRAS.520.2009M}. These programs have observed hundreds of active galactic nuclei, probing a wide range luminosities and redshifts. The OzDES-RM program recently published data  from successful campaigns for eight active galactic nuclei between $0.12 \, < \, z \, < \, 0.71$ \cite{2023MNRAS.520.2009M}. In previous work, they measured the mass of two SMBHs at some of the highest redshifts, $z = 1.905$ and $z =  2.593$.  The measured masses are $4.4^{+2.0}_{-1.9} \times 10^9 M_\sun$ for DES J0228-04 and $3.3^{+1.1}_{-1.2} \times 10^9 M_\sun$ for DES J0033-42, calculated with $f = 4.47 \pm 1.25$. The SDSS-RM Program published results for 849 broad line quasars between $0.1 \, < \, z \, < \, 4.5 $. The observation of an EMRI between the SMBH and a stellar black hole in any candidate observed by these campaigns would result in  multimessenger combinations of reverberation mapping - EMRI, which  constitutes our third benchmark scenario.

\section{Results}
\label{sec:EMRIGW}

In this section, we obtain constraints on the mediator mass versus coupling plane $(m_V, \tilde{\alpha}')$, assuming that the central SMBH and a second stellar black hole constitute an EMRI system. Our strategy will be to compute the temporal evolution of the angular  frequency $\dot{\omega}$, including the correction due to the dark force in Eq.~\eqref{introyovan}, and subsequently compute the  GWs from perturbations of the Kerr metric following \cite{Finn:2000sy} (we refer to \cite{Teukolsky:1973ha, 1978ApJ...225..687D} for classic papers and \cite{PhysRevD.50.6297, PhysRevD.66.044002}  for earlier work). We will then compare our results to the case of pure gravity, where the dark force is absent. Our calculation most closely resembles that of \cite{Alexander:2018qzg}, who performed an analogous calculation for regular equal-mass binary inspirals. We will assume an uncertainty of $\mathcal{O}(10)\%$ in the mass of the SMBH, corresponding to the case where the SMBH mass measurements comes from either stellar kinematics or reverberation mapping. On the other hand, for the case of an XMRI-EMRI multimessenger combination, we will assume that the mass of the SMBH is determined up to negligible fractional uncertainty $\sim \mathcal{O}(10^{-3})\%$.

We note that several prior studies of dark sectors with binary inspirals have concentrated on constraining either the strength of the dipole radiation term, or the charge of the remnant.  For example, \cite{Christiansen:2020pnv, Cardoso:2016olt, Gupta:2021rod, Liu:2020cds, Alexander:2018qzg} consider black hole binary inspirals with massless mediators and constrain the dipole radiation term from the early pre-merger inspiral stage. The charge of the merger remnant can be constrained from the detection of the quasinormal modes. Dipole radiation in the context of EMRIs has been considered by \cite{Liu:2020vsy, Liu:2022wtq, Zhang:2022hbt}. Our focus, as stated before, will instead be on constraining the dark force in the limit of vanishing dipole radiation, relying instead on independent mass measurements of the SMBH.

The use of independent mass measurements to constrain dark forces works under  the assumption that the dark force is only operational between the SMBH and the stellar black hole, while being switched off between the SMBH and surrounding baryonic matter, including stars, other objects such as brown dwarfs, and matter in the accretion disk. This is possible in models where the black holes are charged under a dark force acting only between particles in the dark sector, and not between the dark and visible sectors. We discuss a simple model that achieves this as a proof of principle, in  Appendix~\ref{appdarkcha}. The model consists of millicharged dark fermions with mass $\sim 1$ GeV and charge approximately equal to their mass, near the weak gravity conjecture limit. The dark force carrier is a vector $V$. The discharge time of a black hole initially charged with such dark fermions can be shown to be large under the most conservative assumptions. We remain somewhat agnostic about the question of how the black holes acquire their initial distribution of dark fermions, indicating some possibilities in  Appendix~\ref{appdarkcha}. For the case of the SMBH, the question reverts to the open question of how seeds of such black holes are formed.

\subsection{Effect of Dark Forces}
\label{sec:EMRIGW2}

To set our notation, we consider an EMRI system consisting of two masses $M_1$ and $M_2$, with the mass ratio, $\eta\equiv M_1/M_2 \ll 1$. The heavy object ($M_2$) is a SMBH. In addition to the gravitational attraction between the two objects, we introduce a dark force operating between them. For simplicity, we consider a Yukawa force mediated by a mediator particle $V$
\bea
|\vec{F}_{\rm Yukawa}|=\frac{\alpha' Q_1 Q_2}{r^2} e^{-m_V r} (1+m_V r).
\label{eq:FYukawa}
\eea
Here $\alpha'$ is the dark fine structure constant, $m_V$ is the mass of the dark force mediator, $r$ is the separation between the two objects constituting the EMRI system, and $Q_{i}$ is the dark charge of the object $i$, with $i=1,2$. We will consider both the case when $\alpha' > 0$ (attractive force between charges of opposite signs) as well as $\alpha' < 0$ (repulsive force between charges of the same sign)\footnote{For dark force mediated by a scalar particle, the force between two particles is always attractive. The modifications of inverse-square law from attractive interactions mediated by a scalar mediator and a vector mediator are the same, and we will not distinguish them in the GW analysis.}. Although the Yukawa force is typically used to describe interactions mediator by scalar particles, we do not distinguish between scalar and vector mediators in this study when introducing dark forces in the form of Eq.~\eqref{eq:FYukawa}. The dark force, together with the gravitational attraction, gives the total force acting on each object
\bea \label{modforce2}
|\vec{F}_{\rm total}|=\frac{G M_1 M_2}{r^2}[1+\tilde{\alpha}' e^{-m_V r} (1+m_V r)],
\label{eq:Ftotal}
\eea
where it is convenient to normalize the strength of the dark force  to the gravitational force by introducing
\bea
\tilde{\alpha}'\equiv \frac{\alpha' Q_1 Q_2}{G M_1 M_2}.
\eea
The introduction of the dark force modifies the EMRI system in two ways. Firstly, the angular frequency $\omega(r)$ is determined by the net force on the lighter object, which is modified by the dark force, resulting in a departure from the inverse-square law as given in Eq.~\eqref{modforce2}.  Secondly, dark (dipole) radiation in the form of  emission of the light force mediator will make the inspiral faster, affecting the temporal dependence of the separation $r(t)$. Both $\omega(t)$ and $r(t)$, and therefore  ultimately the GW frequency $\fGW(t)$ and strain $h_c(t)$ are modified from the pure gravity case in the presence of a dark force. In this work, we will assume that the dipole radiation is negligible.

Given a centripetal force in Eq.~\eqref{eq:Ftotal}, the relation between the orbital frequency and the separation $r$ in the presence of a dark force can be obtained as
\bea
\omega^2 = \frac{G (M_1 + M_2)}{r^3} (1 + \tilde{\alpha}' e^{-m_V r} (1 + m_V r) ). 
\label{eq:omegaDelta}
\eea
The GW frequency of a harmonic $m$ is determined with the orbital angular frequency
\bea
f_{{\rm GW},m}=\frac{m}{2 \pi} \, \omega. 
\eea 
Near the ISCO, GW radiation drives orbits to be circular; therefore the $m=2$ mode dominates and we focus on $\fGW\equiv f_{{\rm GW},2}$ in the rest of the paper. 
We determine $r(\omega)$ numerically from Eq.~\eqref{eq:omegaDelta}. The EMRI losses energy mainly through GW radiation when dipole radiation of the mediator is turned off. The GW emission rate is
\bea
\frac{dE_{\rm GW}}{dt}=\frac{32}{5}G \, \mu^2 \, \omega^6 \, r^4.
\eea
Here, $\mu = M_1M_2/(M_1+M_2)$ is the reduced mass.

The time evolution of the orbital frequency due to GW radiation is \cite{Kopp:2018jom}
\bea
\frac{d\omega}{dt} = -\frac{32}{5} G \, \mu \,  \omega^5 \, r^2 \, g \, \mathcal{N}^{-1}.
\label{eq:domegadt}
\eea
The $g$ factor is from the dark force correction to the gravitational attraction,
\bea
g= - \frac{3 + \tilde{\alpha}' e^{- m_V r} \, ( 3 + m_V r (3 + m_V r))}{1 + \tilde{\alpha}' e^{- m_V r} \, (1+ m_V r (1 - m_V r))}.
\label{eq:gfactor}
\eea
While a precise determination of the gravitational wave emission from the EMRI systems remains
an ongoing effort, several sets of waveforms have been provided for data analysis. We adopt the 
result provided by Ref.~\cite{Finn:2000sy}, which was obtained by numerically solving the 
Teukolsky equation~\cite{Teukolsky:1973ha,Sasaki:1981sx} for quasi-circular
orbit cases. The result
for each physical quantity is presented as the usual equation
but with an additional factor added which captures the relativistic
correction. In the case of GW frequency evolution in Eq.~\eqref{eq:domegadt}, we include the correction $\mathcal{N}^{-1}$ obtained from Table VIII of~\cite{Finn:2000sy}.

The strain amplitude of the GW signal can be calculated from the gravitational energy radiation rate. For the characteristic strain in the harmonic $m$, we follow the definition of~\cite{Finn:2000sy},
\bea
h_{c,m} \equiv h_{o,m} \sqrt{2 f_{{\rm GW}, m}^{2} / \dot{f}_{{\rm GW}, m} }\,\,,
\label{eq:hcm}
\eea
with the dominant contribution coming from the harmonic $m=2$. The $f^{2}_{\rm GW} / \dot{f}_{\rm GW}$ factor accounts for the time the GWs are produced at a fixed frequency for detection. The rms amplitude is defined as 
\bea
h_{o,m}= \frac{2 \sqrt{\dot{E}_{\infty m}}}{m \, \omega \, d_L},
\label{eq:hom}
\eea
where $\dot{E}_{\infty m}$ is the total energy emission rate into the $m$-th harmonic to  infinity. $d_L$ is the luminosity distance from the EMRI system to Earth. The full expression for $h_{o,m}$ appears in Eq. (3.11b) of \cite{Finn:2000sy}. We reproduce  the expression for the $m=2$ mode here for completeness
\bea
h_{o,2} = \sqrt{\frac{32}{5}} \, \frac{\eta \, G \, M_2}{d_L} \, (G \, M_2 \, \omega)^{2/3} \, \mathcal{H}_{o,2}.
\label{eq:ho2}
\eea
The relativistic correction to the strain $\mathcal{H}_{o,2}$ can be obtained from the relativistic correction to the energy loss rate $\mathcal{H}_{o,2}=\sqrt{\dot{\mathcal{E}}_{\infty 2}}$, and we use numerical values of $\dot{\mathcal{E}}_{\infty 2}$ reported in Table IV of \cite{Finn:2000sy}. We will focus on the $m=2$ harmonic of the characteristic strain amplitude $h_c\equiv h_{c,2}$, which is calculated through Eq.~\eqref{eq:hcm} with numerical results of $\dot{f}_{\rm GW}$ from Eq.~\eqref{eq:domegadt} and $h_{o,2}$ from Eq.~\eqref{eq:ho2}.

\subsection{Temporal Evolution of Gravitational Waves}

\begin{figure}[h]
  \centering
    \includegraphics[width=0.47\textwidth]{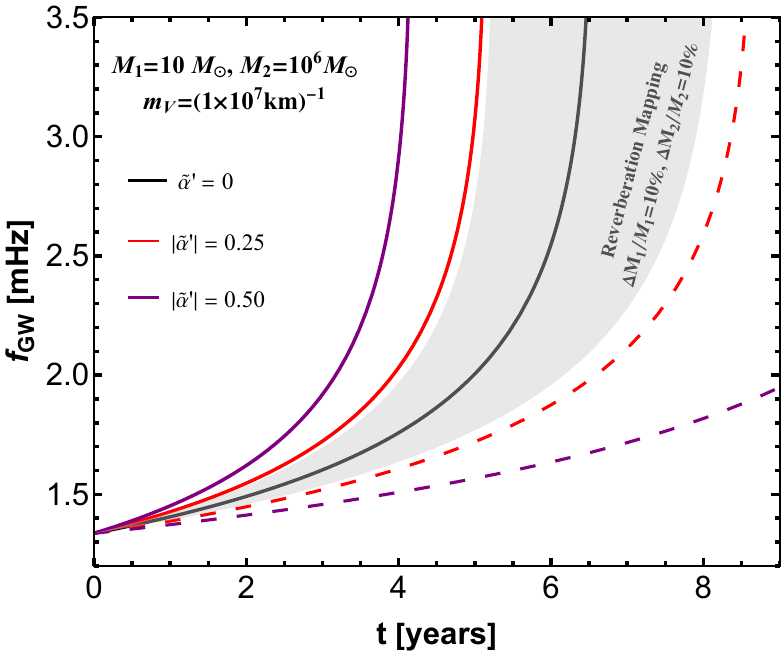} 
    \qquad
    \includegraphics[width=0.47\textwidth]{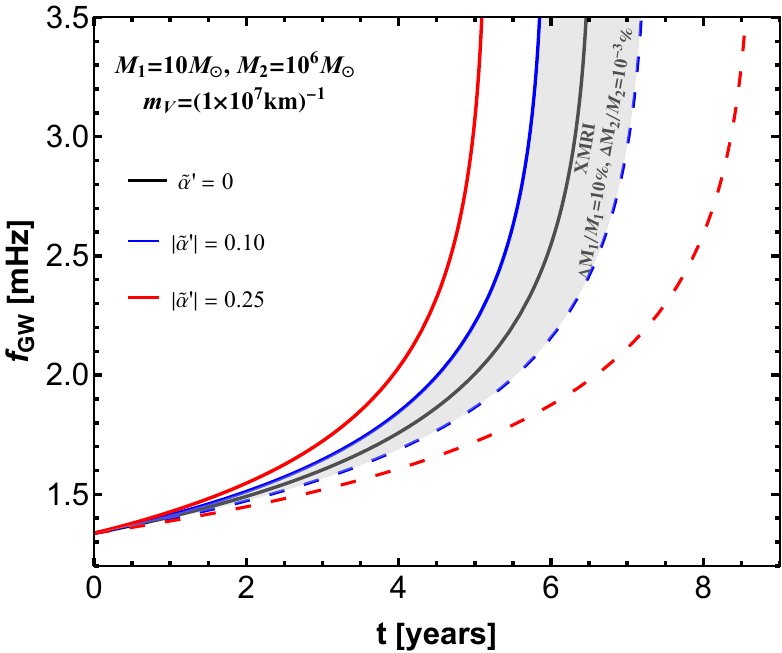} 
\caption{Evolution of the EMRI GW frequency $\fGW$ of a $10 M_{\odot}$ object inspiraling a $10^6 M_{\odot}$ SMBH for a range of dark force strength values $\tilde{\alpha}'$ in the attractive case (solid) and repulsive case (dashed). The mediator mass is $(10^{7}~{\rm km})^{-1}$. (\textit{left}): the case of multimessenger EMRI signals with either reverberation mapping or stellar kinematics data, resulting in an uncertainty (displayed by the gray band) of $10\%$ in determining both of the masses $M_1$ and $M_2$. (\textit{right}): the case of multimessenger EMRI-XMRI signals, resulting in a fractional uncertainty (displayed by the gray band) of  $10^{-3}\%$ in the SMBH mass $M_2$ and $10\%$ in $M_1$.}
\label{fig:fGWevolution}
\end{figure}

\begin{figure}[t]
  \centering
  \includegraphics[width=0.47\textwidth]{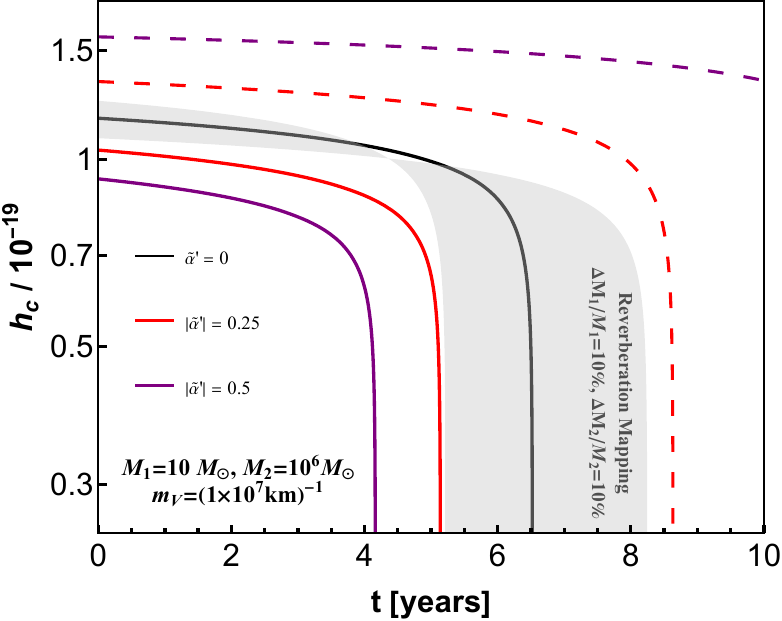} 
  \qquad
  \includegraphics[width=0.47\textwidth]{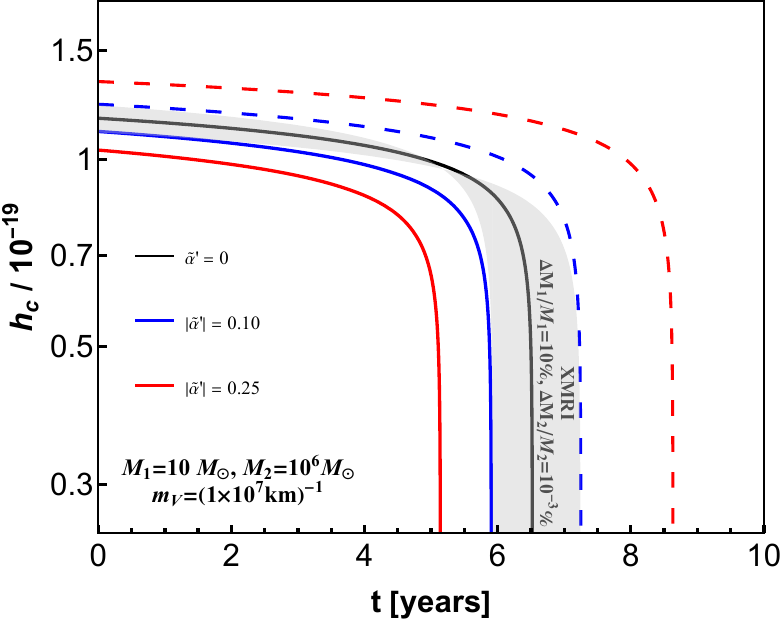} 
  \caption{The temporal dependence of the characteristic strain $h_{c}$ of the EMRI system for a range of dark force strength values in the attractive (solid) and repulsive (dashed) cases. The black hole and dark force mediator masses are chosen to be consistent with Fig.~\ref{fig:fGWevolution}. (\textit{left}): multimessenger measurement with reverberation mapping. The uncertainty of masses $M_1$ and $M_2$ are both $10\%$. (\textit{right}): multimessenger measurement with XMRI. The uncertainty of masses $M_1$ and $M_2$ are $10\%$ and $10^{-3}\%$ respectively.}
\label{fig:hcevolution}
\end{figure}

We use formulas introduced in the last section to calculate GWs emitted by the EMRI system. Fig.~\ref{fig:fGWevolution} and Fig.~\ref{fig:hcevolution} illustrate the evolution of the GW frequency $f_{\rm GW}$ and characteristic strain $h_c$ for different choices of the dark force strength parameter $\tilde{\alpha}'$. The mass of the SMBH is chosen as $M_2=10^{6} M_{\sun}$, and the mass of the inspiralling black hole is $M_1=10 M_\sun$. We assume for simplicity both black holes are Schwarzschild black holes, though dark forces can also impact EMRIs involving rotating black holes. The mass of the mediator is $m_V = (10^{7}~{\rm km})^{-1}\simeq 1.98\times 10^{-17}{\rm eV}$, such that the force range is comparable to the ISCO radius of $8.9\times10^{6}~{\rm km}$ for a $10^6 M_\odot$ SMBH. The associated GW frequency emitted at the ISCO is about $4.4~{\rm mHz}$. We evolve the EMRI system using Eq.~\eqref{eq:domegadt} from the initial condition $f_{\rm GW}=1.34~{\rm mHz}$ at $t=0$, and calculate the GW emission power with Eq.~\eqref{eq:hcm}. The corresponding initial EMRI components separation is about $2\times10^7~{\rm km}$ with small difference depending on the type of interactions.

The colored curves in Fig.~\ref{fig:fGWevolution} show $\fGW$ as a function of time in the cases with the absolute value of the dark force strength $|\tilde{\alpha}'|=0.5$ (purple), $0.25$ (red), and $0.1$ (blue). The attractive force case ($\tilde{\alpha}'>0$) is shown with solid curves while the repulsive force case ($\tilde{\alpha}'<0$) is shown with dashed curves. As a comparison, we show the case in the absence of dark force with the black curve, together with a gray band representing the typical uncertainty from the determination of the black hole masses. In the left panel, we show the uncertainty of $\Delta M_2 / M_2=10\%$ corresponding to reverberation mapping campaigns. In the right panel, the uncertainty is $\Delta M_2 / M_2=10^{-3}\%$ corresponding to the mass determination with an XMRI measurement. In both panels, a $10\%$ uncertainty for the mass of the inspiraling black hole $\Delta M_1/M_1=10\%$ is incorporated within the gray bands. Evidently, the attractive dark force accelerates the system towards ISCO, while the repulsive dark force slows down the system's evolution towards ISCO. Larger values of the effective dark force strength induce more significant deviations from the gravitation-only scenario. With a more precise multimessenger determination of $M_2$ in the right panel, a smaller $|\tilde{\alpha}'|=0.1$ can be distinguished from the uncertainty band compared to the sensitivity of about $|\tilde{\alpha}'|=0.25$ in the left panel.

In Fig.~\ref{fig:hcevolution}, we show the strain amplitude $h_{c}$ as a function of time for the same EMRI event consisting $M_1=10 M_\odot$ and $M_2=10^{6} M_\odot$ located at $d_L=1~{\rm Gpc}$, in the attractive dark force case (solid) and the repulsive dark force case (dashed). The inverse mediator mass is chosen to be $1\times10^{7}~{\rm km}$. In all figures, the pure gravity case is shown with the black curve, and the benchmarks with non-zero dark force couplings are shown for $|\tilde{\alpha}'|=0.5$ (purple), $|\tilde{\alpha}'|=0.25$ (red), and $|\tilde{\alpha}'|=0.1$ (blue), corresponding to the same benchmark values in Fig.~\ref{fig:fGWevolution}. The SMBH mass uncertainty is shown with the gray band for $10\%$ in the left panel, and for $10^{-3}\%$ in the right panel, while the mass uncertainty of the lighter component is kept to be $10\%$ in both panels. We note that for an attractive (repulsive) dark force, the characteristic strain $h_{c}$ is weaker (stronger) than the pure gravity case at a given time $t>0$. This contrasts with the evolution of rms amplitude $h_{o,2}$ of GWs emitted toward infinity. The dependence is reversed as a result of the effect of the dark force on the GW frequency: when the EMRI system evolves faster to ISCO, it spends less time $\fGW/\dot{f}_{\rm GW}$ in the fixed observation frequency window, such that the $h_{c}$ in Eq~\eqref{eq:hcm} is suppressed. Similar to Fig.~\ref{fig:fGWevolution}, we see that the effect of the dark force can be discerned best with a stronger dark force strength $|\tilde{\alpha}'|$.

The strain amplitude of a resolved EMRI signal is inversely proportional to the luminosity distance. Therefore, EMRI events that are closer can be measured with greater precision; indeed, as mentioned before, multimessenger studies involving EMRIs rely on stellar kinematics or reverberation mapping that are only effective at much smaller distances. Nevertheless, $d_L = 1 ~{\rm Gpc}$ represents a conservative choice.

\begin{figure}[t]
  \centering
\includegraphics[width=0.47\linewidth]{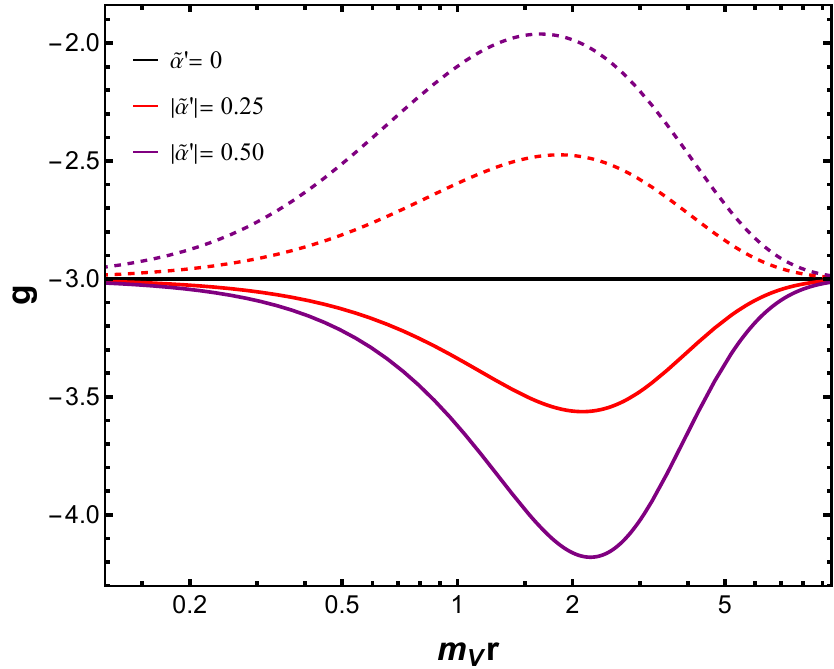}
  \caption{The coupling $g$ defined in Eq.~\eqref{eq:gfactor} as a function of the dimensionless force range $m_V r$ with different choices of $\tilde{\alpha}'$ for attractive dark force (solid) and repulsive dark force (dashed). The largest deviations from the gravitation only case, $g=-3$, occur when $m_V r \sim 1$. }
  \label{fig:gfactor}
\end{figure}

In addition to the dark force strength, EMRIs are also sensitive to the mass of the mediator $m_V$, which determines the range of the dark force. The dependence on the mediator mass is from the $g$ factor in Eq.~\eqref{eq:gfactor}. In the scenario where there is no dark force, as well as in the scenario where the range of the dark force is much longer or shorter than the EMRI separation, Eq.~\eqref{eq:gfactor} reduces to $g=-3$. Conversely, the correction factor deviates from $-3$ when $m_V$ is comparable to the inverse mutual separation of the EMRI components. We show the dependence of the $g$ factor on the relative values of $m_V$ and $r$ in Fig.~\ref{fig:gfactor}, for the case of attractive (denoted by solid lines) as well as repulsive (denoted by dashed lines) dark force interactions. It is clear that significant deviations from the gravity only case occur when $m_V r\sim 1$. The deviation of the $g$ from $-3$ introduces an effect on the EMRI evolution that agrees with the attractive or repulsive nature of the dark force. Specifically, the system evolves more quickly for an attractive dark force interaction when $g$ is more negative around $m_V r \sim 1$, while it evolves more slowly for a repulsive dark force interaction when $g$ is less negative around $m_V r \sim 1$.

\begin{figure}[t]
  \centering
    \includegraphics[width=0.45\textwidth]{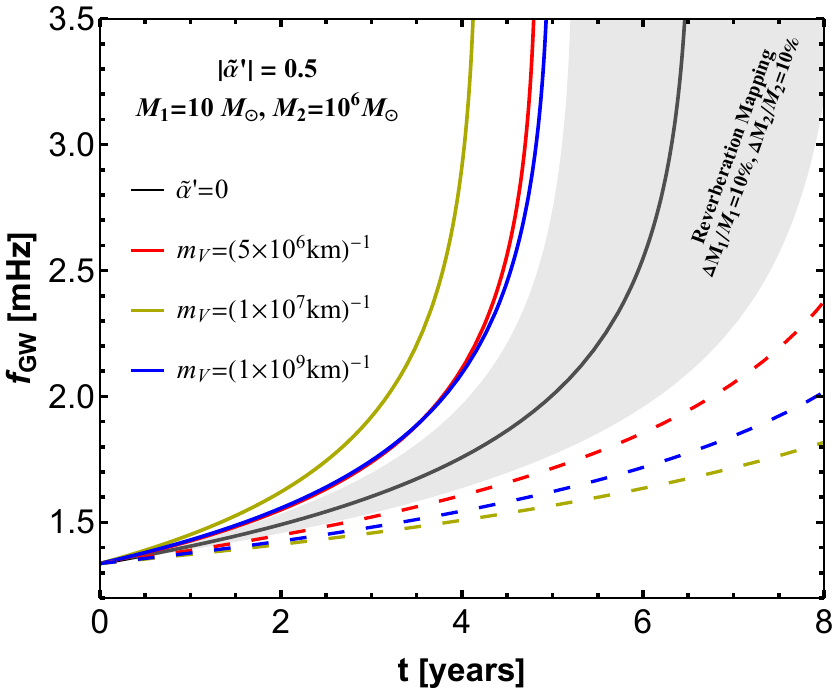} 
\quad
    \includegraphics[width=0.45\textwidth]{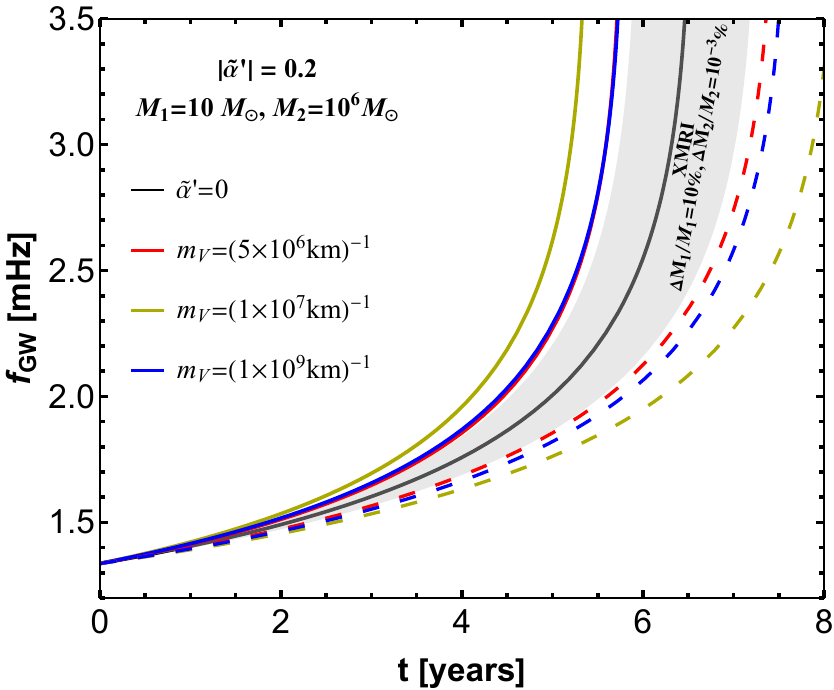} 
\caption{GW frequency versus time plot of a $10 M_{\odot}$ object inspiraling a $10^6 M_{\odot}$ SMBH for a range of the dark force mediator mass $m_V$. (\textit{left}): the case of multimessenger EMRI signals with reverberation mapping data, resulting in an uncertainty of $10\%$in determining each of the masses $M_1$ and $M_2$. (\textit{right}): the case of multimessenger EMRI-XMRI signals, resulting in an uncertainty of  $10^{-3}\%$ in the SMBH mass $M_2$ and $10\%$in $M_1$.}
\label{fig:mediators}
\end{figure}

The effect of the mediator mass on the EMRI is displayed in Fig.~\ref{fig:mediators} with $f_{\rm GW}$ as a function of time for a fixed dark force strength but different mediator masses. We choose $\tilde{\alpha}'=0.5$ in the left panel, and  $\tilde{\alpha}'=0.2$ for the right panel. The other EMRI parameters are kept to be consistent with Fig.~\ref{fig:fGWevolution}. We also choose the initial frequency to be $f_{\rm GW}=1.34~{\rm mHz}$ for all EMRIs. The corresponding initial black hole separation is about $2\times10^7~{\rm km}$ with small difference depending on the type of interactions. As a comparison, the ISCO radius is about $8.9\times10^{6}~{\rm km}$ for $M_2=10^{6} M_\odot$. The mediator mass values are chosen to represent three different force range scenarios: short force range $m_V^{-1}=5\times10^6~{\rm km}$ (red); intermediate force range comparable to the ISCO radius $m_V^{-1}=10^7~{\rm km}\sim R_{\rm ISCO}$ (yellow); and the very long-range force limit $m_V^{-1}=10^9~{\rm km}$ (blue). As is expected from the dependence of the $g$ factor on $m_V\, r$, the dark force effect of yellow curves exhibit more significant distinction from the gravity-only case (black), compared to the short-range and long-range force limits.

In the regime where $m_V \ll R_{\rm ISCO}^{-1}$, EMRI GWs can still discern between the gravitational attraction and an additional dark force if one advocates, as we have, for an independent mass measurement for which the dark force is switched off. This is different from the case of GW measurements from equal-mass binary systems, where the effect of the dark force becomes entirely degenerate with gravity: in that case, a rescaling of the binary component mass by a factor of $(1+\tilde{\alpha}')^{2/5}$ mimics the presence of the dark force. The primary challenge in distinguishing the effect of the dark force stems from uncertainties in the pure gravitational contribution, which ultimately arises due to uncertainties in $M_1$ and $M_2$ (illustrated by the gray band). For instance, for a benchmark value of $\tilde{\alpha}'=0.5$, the rescaling factor on the mass, $(1+\tilde{\alpha}')^{2/5}\simeq 1.18$, exceeds the uncertainty in the reverberation mapping and the XMRI measurements. Consequently, the degeneracy is broken for EMRI GWs in the ultra-long-range force limit.

\begin{figure}[t]
  \centering
    \includegraphics[width=0.46\textwidth]{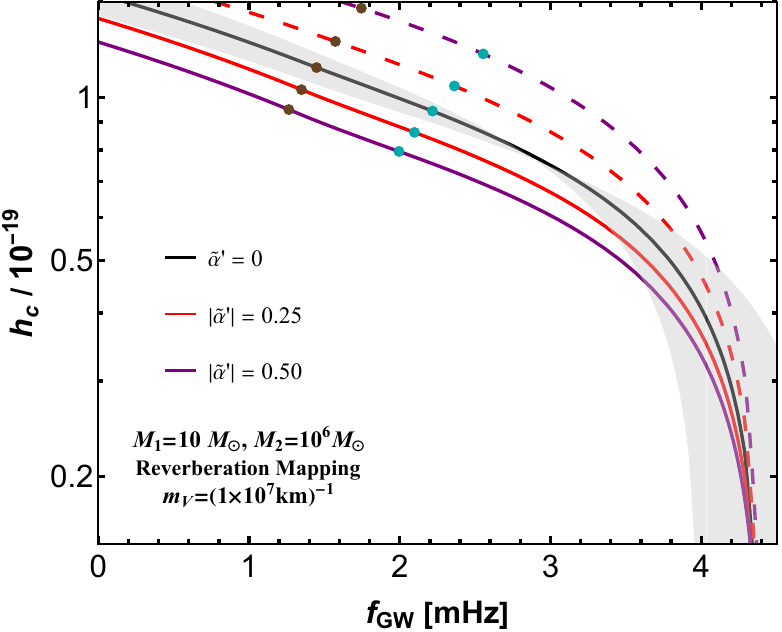}
    \qquad
    \includegraphics[width=0.47\textwidth]{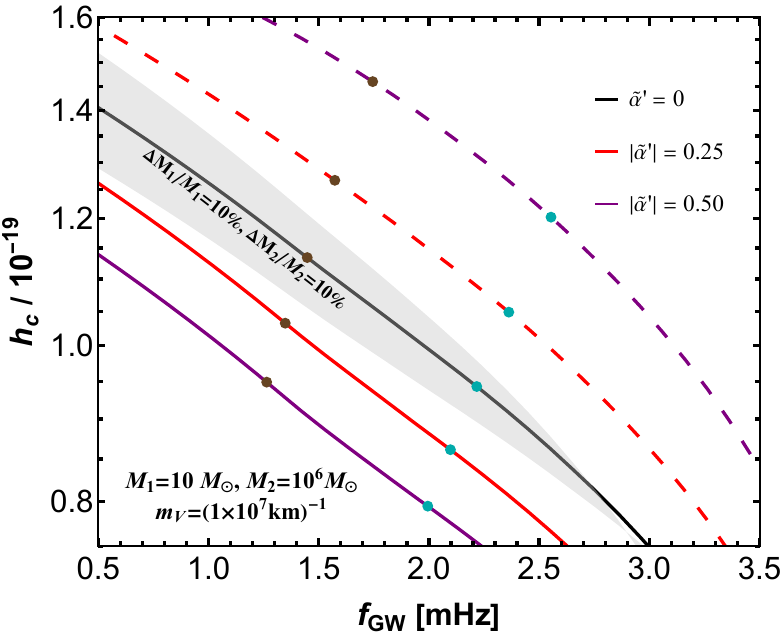} \\
    
    \vspace{7pt}
    \includegraphics[width=0.46\textwidth]{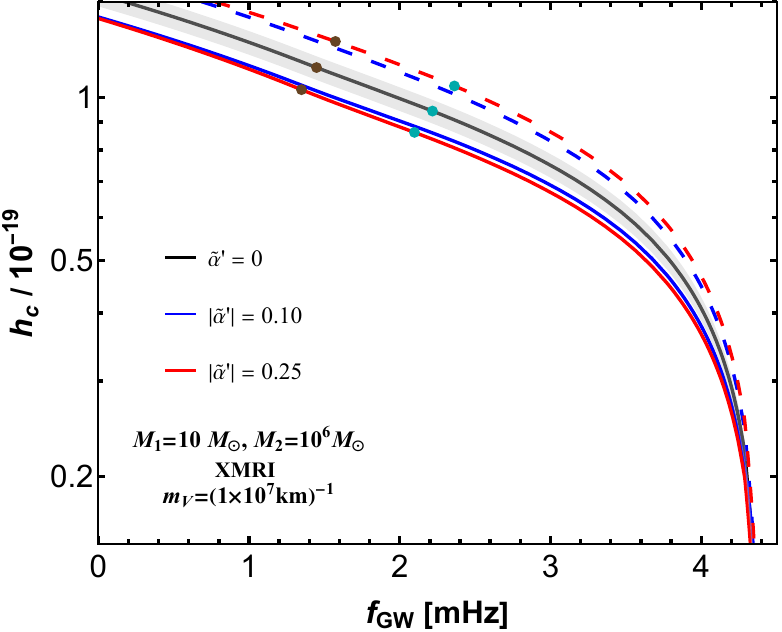} 
    \qquad
    \includegraphics[width=0.47\textwidth]{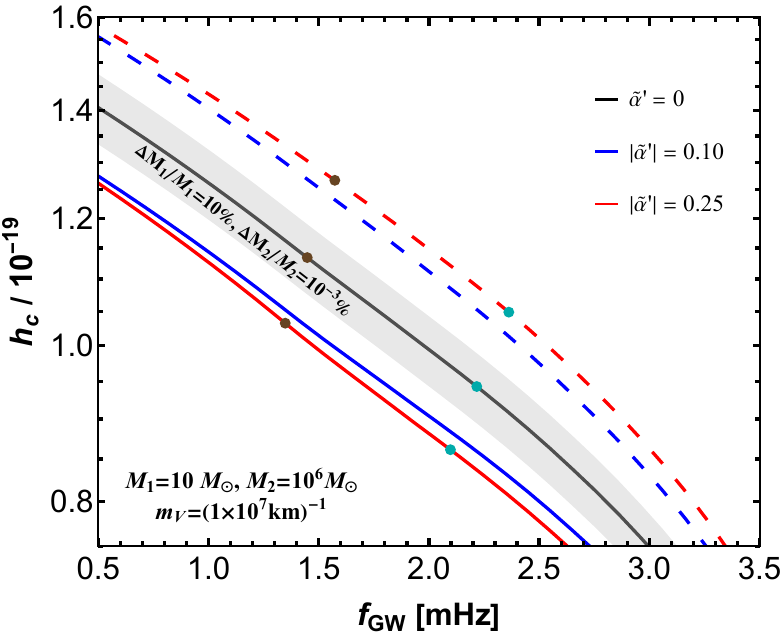}
    
\caption{Strain $h_{c}$ as a function of the GW frequency $\fGW$ near ISCO. Color curves show the EMRI evolution with a range of dark force strengths. Black curve shows the evolution in a gravitation-only scenario. Brown dots indicate the evolution time is 5 years from ISCO while cyan dots represents 1 year left to ISCO. (\textit{upper}): Dark force effect compared to the gravitation-only case, with mass uncertainties $\Delta M_1 / M_1 = 10\%$ and $\Delta M_2 / M_2 = 10\%$, depicted on the left, and zoom-in of specific regions on the right. (\textit{lower}): Dark force effect compared to the gravitation-only case with $\Delta M_1 / M_1 = 10\%$ and $\Delta M_2 / M_2 = 10^{-3}\%$ on the left, and zoom-in plot on the right.}
\label{fig:fGWhc2}
\end{figure}

In the end, we show the EMRI GWs on the plane of strain versus frequency in Fig.~\ref{fig:fGWhc2}, with the black and cyan dots indicating the remaining time from ISCO is $5~{\rm years}$ and $1~{\rm year}$ respectively. The mediator mass is set to $m_V^{-1}=10^7~{\rm km}$ to demonstrate the maximum effect. The upper panels illustrate the effects of the dark force compared to uncertainties determined by reverberation mapping, with a detailed zoom-in on the region where the dark force leads to a significant deviation from the gravity-only scenario. The lower panels show the case when uncertainties are determined by the XMRI measurement. The GW data collected over the last few years from the EMRI events in future observations will be utilized to precisely track the evolution of the strain versus the frequency, crucial for distinguishing the existence of the dark force. We will use the $h_c-\fGW$ relation to calculate the sensitivity to the dark force in the next section.

\subsection{Sensitivity with Multimessenger Measurements}

We perform a likelihood analysis to test the future sensitivity to $\tilde{\alpha}'$ with the EMRI GWs and the additional independent measurement of the black hole masses. The likelihood analysis is based on the strain-frequency data of the GW signal. The likelihood is defined as \cite{Maggiore:2007ulw}
\begin{equation}
    \Lambda (s|\tilde{\alpha}') = \mathcal{K} \exp [(h_{\tilde{\alpha}'}|s) - \frac{1}{2}(h_{\tilde{\alpha}'}|h_{\tilde{\alpha}'}) - \frac{1}{2}(s|s) ]
    \label{eq:likelihood}
\end{equation}
Here $\mathcal{K}$ is a normalization constant, $s(t)$ is the observed GW waveform in the time domain, and $h_{\tilde{\alpha}'}(t)$ is the waveform corresponding to a selected value of $\tilde{\alpha}'$. The scalar product between the two waveform is given by

\begin{equation}
    (h_{\tilde{\alpha}'}|s) = 4 \, {\rm Re} \int_{f_{\rm min}}^{f_{\rm max}} d \fGW  \,\frac{\tilde{h}_{\tilde{\alpha}'}(\fGW) \, \tilde{s}(\fGW)}{S_n(\fGW)}
\end{equation}
where $S_n(\fGW)$ represents the noise spectral density of the GW observatory and has a dimension of $\text{Hz}^{-1}$. Quantities denoted with a tilde represent the Fourier transform of the time domain data. Following the definitions in~\cite{Moore:2014lga}, we can write the scalar product as 
\begin{equation}
    (h_{\tilde{\alpha}'}|s) =  \, {\rm Re} \int_{f_{\rm min}}^{f_{\rm max}} d\fGW \, \frac{h_{c , \tilde{\alpha}'}(\fGW) s_c(\fGW)}{\fGW^2 \, S_n(\fGW)}
    \label{eq:scalarproduct}
\end{equation}
The characteristic strain $h_{c,\tilde{\alpha}'}(\fGW)$ is calculated with the chosen dark force strength. The $s_c(\fGW)$ is obtained from the observed GW waveform. In the EMRI measurement, the upper limit $f_{\rm max}$ and the lower limit $f_{\rm min}$ of the integral in Eq.~\eqref{eq:scalarproduct} is determined by the frequency range over which the waveform is observed, with the upper limit set to $f_{\rm GW}^{\rm ISCO}$ and the lower limit set to the EMRI GW frequency at $10$ years before ISCO \footnote{The LISA observatory is set to observe for $4.5$ years, with the possibility of extension up to 10 years \cite{LISA:2017pwj, Colpi:2024xhw}. For the proposed $\mu \text{Ares}$ mission, a 10-year-long observation period is foreseen. Therefore, we use $t_{\rm obs} = 10~{\rm years}$ in our analysis.}. As long as the EMRI signal is stronger than the detector noise, the likelihood is mostly determined by the deviation of the signal spectrum in the presence of the dark force from the gravity-only spectrum (with mass uncertainties). We calculate the value of $\tilde{\alpha}'$ for a given $m_V$ that maximizes the likelihood. This is equivalent to maximizing the log-likelihood, which we can read from Eq.~\eqref{eq:likelihood} as
\begin{align}
    \log \, \Lambda(s|\tilde{\alpha}') & = \log \, \mathcal{K} +  (h_{\tilde{\alpha}'}|s) - \frac{1}{2}(h_{\tilde{\alpha}'}|h_{\tilde{\alpha}'}) - \frac{1}{2}(s|s) \nonumber\\
    & = (h_{\tilde{\alpha}'}|s) - \frac{1}{2}(h_{\tilde{\alpha}'}|h_{\tilde{\alpha}'})
\end{align}
For an observed waveform $s(t)$, the term $\text{log} \, \mathcal{K} - \frac{1}{2}(s|s) $ is a constant independent of the dark force effect and can be omitted from the definition of log-likelihood in the rest of the analysis.
 
To estimate the sensitivity to $\tilde{\alpha}'$, we assume that the observed signal corresponds to the dark force scenario. The likelihood of a gravity-only scenario is calculated with the assumed uncertainties in the masses of the EMRI components and $\tilde{\alpha}' = 0$. A scan across the parameter space of $\tilde{\alpha}'$ is performed to assess the likelihood with non-vanishing $\tilde{\alpha}'$. We further assume that the test statistic, denoted as $\tau$, is $\chi^2$-distributed under the null hypothesis. The two likelihoods are then compared at a $95 \%$ confidence level. The likelihood test statistics is
\begin{equation}
    \tau =  -2 \, \left( \ln \, \Lambda(s|0) - \max_{{\tilde{\alpha}}\geq0} \left[ \ln \, \Lambda(s|\tilde{\alpha}) \right] \right)
    \label{eq:loglike}
\end{equation} 

We will use the likelihood test statistics to calculate the sensitivity to the dark force strength and the force range. While the precision of GW measurements improves the sensitivity to $\tilde{\alpha}'$, the sensitivity to the dark force range depends on the mass of the SMBH involved in the EMRI event. We discuss the relation between the dark force range and the SMBH mass in the following. EMRI GWs are observed mostly when the orbits of the inspiraling object have decayed close to the ISCO radius, which can be written as $R_{\rm ISCO}= 6 \, G M_2$ for Schwarzschild SMBHs. The frequency of GWs emitted when the separation is equal to the ISCO radius, $r=R_{\rm ISCO}$, is
\bea
f_{\rm GW}^{\rm ISCO} = \frac{1}{6\sqrt{6}\,\pi\,G\, M_2}\simeq 4.35 \, \left(\frac{10^6 M_{\odot}}{M_2}\right)~{\rm mHz}.
\eea
This determines the observed GW frequency for EMRIs with different center SMBH masses. The frequency window of LISA is ideal to be used for observing GWs from $\sim10^{6}M_{\odot}$ SMBHs. On the other hand, EMRI GWs from heavier SMBHs appear at lower frequencies that are outside the LISA range. For example when $M_2\sim10^{9} M_{\odot}$, the GW frequency lies in the unexplored $\mu$Hz range. Future observations of $\mu$Hz GWs have been proposed with space-based interferometers $\mu$Ares~\cite{Sesana:2019vho}, cold-atom interferometers~\cite{AEDGE:2019nxb}, and using Solar System asteroids as test masses~\cite{Fedderke:2021kuy}. With a larger SMBH mass and consequently a wider ISCO radius, we can probe the dark force of a lighter mediator mass $m_V$, which would be indistinguishable from a rescaling of gravitational force with measurements at smaller separations. Therefore, the mediator mass of optimal sensitivity, the SMBH mass and the GW frequency can be correlated as follows,
\bea
m_V^{-1} \simeq 8.96 \times 10^{8} \, \left(\frac{M_2}{10^{8} M_{\odot}}\right)~{\rm km} \simeq 3.90 \times 10^{8} \,\left(\frac{10^{-4}~{\rm Hz}}{f_{\rm GW}}\right)~{\rm km} 
\label{eq:risco8}
\eea

\begin{figure}[t]
  \centering
    \includegraphics[width=0.5\textwidth]{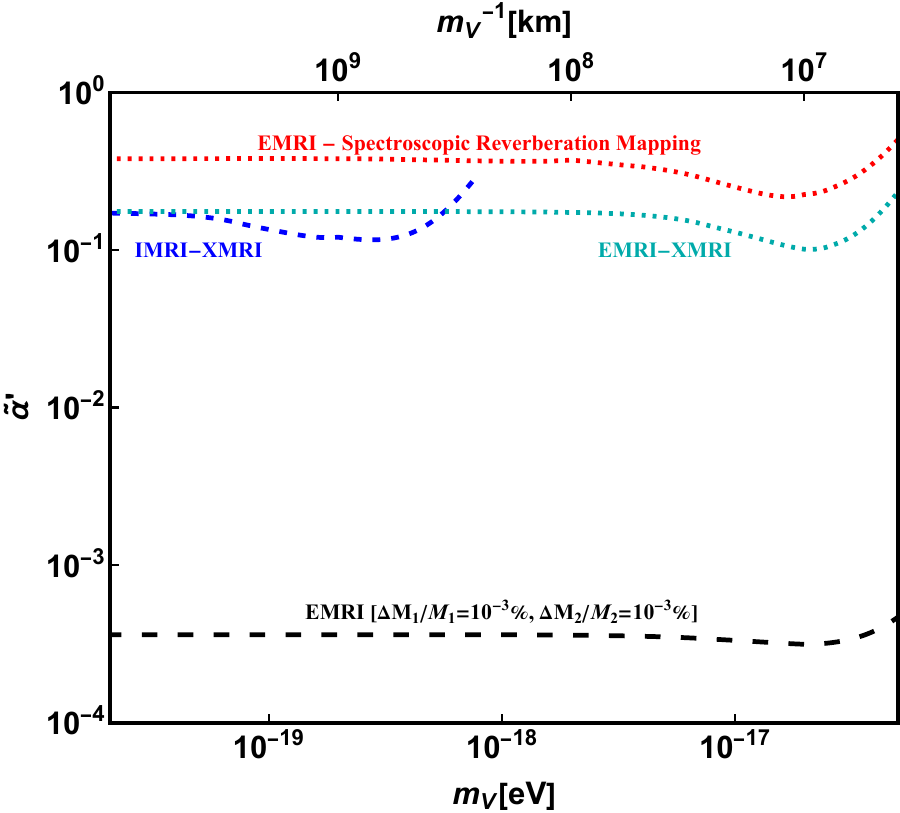}    
    
    \caption{Projected $95\%$ C.L. sensitivity of LISA and $\mu$Ares to an attractive dark force on its strength $\tilde{\alpha}'$ and force mediator mass $m_V$. The cyan, red, and black curves assume the EMRI system consists of a stellar black hole of mass $M_1=10M_{\odot}$ and a SMBH of mass $M_2=10^6M_{\odot}$. The cyan dotted curve corresponds to the case of EMRI-XMRI signals, resulting in an uncertainty of $10\%$ in determining the value of $M_1$, and $10^{-3}$\% in determining the value of $M_2$. The red dotted curve displays the sensitivity of EMRI signals complemented with spectroscopic reverberation mapping measurements, leading to a $10\%$ uncertainty in both $M_1$ and $M_2$. The black dashed curve depicts the sensitivity of an EMRI waveform observed with an uncertainty of $10^{-3}$\% in determining both masses. We also show the case of a heavier SMBH mass with the blue dashed curve for an IMRI-XMRI multimessenger measurement. The black hole masses are $M_1=10^{4}M_{\odot}$ and $M_2=10^8M_{\odot}$, with $10\%$ uncertainty in $M_1$ and $10^{-3}\%$ uncertainty in $M_2$.}
  \label{fig:alphaM}
\end{figure}

We present the sensitivity to $\tilde{\alpha}'$ across a range of the mediator mass $m_V$ in Fig.~\ref{fig:alphaM}. In particular, we choose two sets of representative GW event parameters: (1) $M_2=10^6M_\odot, d_L=10~{\rm Mpc}$ for lighter SMBHs; and (2) $M_2=10^8M_\odot, d_L=0.1~{\rm Gpc}$ for heavier SMBHs, to demonstrate their sensitivity to different mediator masses. In both benchmarks, the results are calculated by requiring $\tau = 4$ in Eq.~\eqref{eq:loglike} with different multimessenger measurements. In both cases, we will show that the optimal sensitivity is achieved when the mediator mass approximates the inverse of the EMRI separation, as can be seen from the minimum of the sensitivity curve at $m_V r \sim 1$.

In the first benchmark, the SMBH mass uncertainty can be independently constrained to within $10\%$ by spectroscopic reverberation mapping or stellar kinematics, with which the sensitivity of EMRI GW measurement is shown with the red dotted curve in Fig.~\ref{fig:alphaM}. Additionally, the SMBH mass can be measured within $10^{-3}\%$ uncertainty with XMRIs. The corresponding sensitivity is shown with the cyan dotted curve in Fig.~\ref{fig:alphaM}. The EMRI GWs from $\sim10^6M_\odot$ SMBHs will be measured precisely with the future LISA observatory. We use noise power spectral densities reported in \cite{PhysRevD.95.103012} for the LISA instrumental noise and the galactic confusion noise to obtain the sensitivity. We assume $M_1=10M_\odot$ with $10\%$ uncertainty. The results show that LISA has optimized sensitivity to $m_V\sim 10^{-17}-10^{-16}~{\rm eV}$, while it can still detect effects from longer mediator wavelengths. To illustrate sensitivities achieved through further reduction of uncertainties associated to the black hole masses, we present the sensitivity corresponding to $10^{-3}$\% precision for both $M_1$ and $M_2$ in the black dashed curve.

In the second benchmark, we assume the sensitivity is set through GWs measured by the future $\mu$Ares observation, combined with an independent SMBH mass measurement $\Delta M_2/M_2=10^{-3}\%$ from the XMRI. $\mu$Ares is capable of detecting GWs with a sensitivity of $h_c \approx 10^{-18}$ at $\fGW = 10^{-6}~{\rm Hz}$. With such sensitivities, it will be able to observe Intermediate Mass Ratio Inspirals (IMRIs) involving $10^3 - 10^4$ $M_\odot$ intermediate-mass black holes inspiralling onto $10^7 - 10^8$ $M_\odot$ SMBHs out to $z\geq6$ \cite{Sesana:2019vho}. An IMRI system will respond to the dark force similarly to an EMRI, and the GW emission will be altered in a comparable manner. According to Eq.~\eqref{eq:risco8}, an IMRI signal from a $10^8 M_\odot$ SMBH is able to probe $m_V^{-1} \gtrsim 10^{9}~{\rm km}$. We assume $M_1= 10^4 M_\odot$ for the intermediate-mass black hole, and use the $\mu$Ares noise level reported in~\cite{Sesana:2019vho} to calculate its sensitivity. The final sensitivity is depicted by the blue dashed curve in Fig.~\ref{fig:alphaM}. In the regions where the blue curve overlaps with the red and cyan curves, the dark force effect can be measured for both SMBHs of different masses, allowing the parametric dependence on $\tilde{\alpha}'$ and $m_V$ to be disentangled.

Our study concentrates on the role of independent mass measurements; however, previous studies employing Fisher information matrix analysis demonstrated that uncertainties in EMRI parameters could potentially be reduced to an order of magnitude of $\mathcal{O}(10^{-4})$~\cite{Barack:2003fp}, benefiting from the information in the GW waveform (see also applications to dark force effects in neutron star binaries~\cite{Alexander:2018qzg}). In order to estimate the prospective sensitivity of an EMRI observation to the strength of the dark force utilizing the maximum likelihood criterion, we use the mass determination uncertainty of $10^{-3}\%$ to demonstrate the improvement of the sensitivity with both the multimessenger measurements and the dedicated waveform analysis. While a comprehensive waveform analysis that takes into account the additional dark force coupling within the model parameters is outside the scope of this study, our findings indicate that an independent measurement of the SMBH mass—one not influenced by the dark force— could resolve the degeneracy between the EMRI model parameters.

\section{Conclusions}
\label{sec:conclusion}

Our focus in this paper has been to probe dark forces using EMRI systems. In contrast to other studies using compact object mergers, we have stressed on the importance of independent mass measurements of the central SMBH that do not depend on the interaction between the SMBH and stellar black hole partner of the EMRI system. Such independent mass measurements  break the degeneracy of the effect of the dark force with a simple rescaling of the binary component masses. We have discussed two classes of such independent mass measurements, based on multimessenger studies involving other gravitationally bound systems and on photons.  In the first class are XMRI systems involving the SMBH and a smaller object, such as a brown dwarf. In the second class are various existing methods of determining the SMBH mass, such as stellar kinematics and reverberation mapping campaigns. The SMBH mass determination in the first case can be very precise, with an uncertainty of $\sim 10^{-3}\%$, while a standard benchmark of the mass uncertainty using reverberation mapping is $\sim 10\%$. We have depicted both benchmarks in our results.

The question of localization of EMRIs using future GW detectors is important for our work. The LISA-TianQin or TaiJi-TianQin networks have the possibility of localizing GW events; such localization is frequency-dependent and has been estimated to be $200$, $3$, and $0.005$ $\text{deg}^2$ at $1$, $10$, and $100$ mHz, respectively \cite{Zhang_2021}. A typical benchmark for the localization of EMRIs is $\lesssim 10$ $\text{deg}^2$ \cite{Berry:2019wgg}.  Multimessenger observations involving electromagnetic follow-ups, with the accretion disk or the broad line region, would further localize EMRIs (their potential as standard candles has been discussed  by \cite{LIGOScientific:2017adf}). 

In this study, we have assumed that the corrections from environmental effects are negligible. SMBHs are known to exist in dense astrophysical environments, which may influence the dynamics of EMRIs~\cite{Barausse:2014tra, Berti:2015itd, Kejriwal:2023djc}. Key processes such as gravitational pull from matter, accretion onto EMRI components, dynamical friction, and planetary migration can impact the gravitational waveform, potentially leading to biased parameter estimation and reduced detectability of EMRIs. We direct readers to~\cite{Barausse:2014pra} for information on the maximum level of corrections due to various environmental effects over a typical LISA mission. The environmental corrections are expected to be minimal for SMBHs surrounded by thick disks, but more substantial for those in thin disks, which are predominantly found in active galactic nuclei at high redshifts. Although SMBHs in thin disks contribute up to only a few percent of the EMRIs detectable by LISA with $z\lesssim1$, potential uncertainties from environmental effects should also be considered when applying our method to these events.

Our results motivate multimessenger studies with coordination between GW detectors and reverberation mapping campaigns. In particular, SMBH mass determination by the OzDES-RM and SDSS-RM programs, and the AGN Black Hole Mass Database serve as rich targets for future EMRI studies. The upcoming SDSS-V Black Hole Mapper Reverberation Mapping Project is also important in this regard \cite{Fries_2023}. Similarly, SMBH mass determination by the SINFONI and MASSIVE Surveys using dynamical tracers also serve as possible targets for future EMRI sources. The best case scenario is the concurrent observation of an EMRI and an XMRI for a given SMBH.

\acknowledgments
The authors would like to thank Huai-Ke Guo  and Chen Sun for collaboration in the early stages of the project. KS would like to thank the Aspen Center for Physics for hospitality during the course of this work. TX would like to thank the Center for Theoretical Underground Physics and Related Areas (CETUP*), The Institute for Underground Science at Sanford Underground Research Facility (SURF), and the South Dakota Science and Technology Authority for hospitality during a period part of this work was completed. The work of KS and TX is supported by the U.S.~Department of Energy Grant DE-SC0009956.

\appendix

\section{Dark Charged Black Holes} \label{appdarkcha}

In this appendix, we discuss the existence of black holes with dark charge. Several mechanisms of endowing black holes with long-lived charge content are mentioned by the authors of \cite{Bozzola:2020mjx, Dror:2021wrl} - one interesting possibility is to introduce millicharged fermions in the dark sector \cite{Cardoso:2016olt}. Denoting the mass of the SMBH by $M_2$ and its charge by $Q_2$, one can obtain $Q_2/M_2 \sim 1$ by choosing the charge-to-mass ratio of dark fermions to be $\sim 1$ (close to the limit imposed by the weak gravity conjecture).  In this case, one could consider a SMBH-brown dwarf XMRI system (where the brown dwarf is the uncharged Object A alluded to above) which determines the mass of the SMBH; concurrently, a SMBH-black hole EMRI system (where the stellar black hole is the dark-charged Object B) constrains the light mediator mass. We outline the model below.

Our model will consist of a millicharged dark fermion $\chi$ with charge $q_\chi$ and mass $m_\chi$, interacting by the exchange of a vector mediator $V_\mu$. The Lagrangian is 
\be
\mathcal{L} \, = \, -\frac{1}{4}V_{\mu\nu}V^{\mu\nu} + \frac{1}{2}m_V V_\mu V^\mu + \bar{\chi}(i\gamma_\mu D^\mu - m_\chi) \chi \,\,.
\ee
Here, the covariant derivative is given by $D_\mu = \nabla_\mu + i q_\chi V_\mu$, and we normalize the dark fine structure constant $\alpha^\prime = 1$\footnote{For comparison, \cite{Kopp:2018jom} sets $q_\chi = 1$.}. The total charge of the SMBH can then be written as 
\be
Q_2 \, = \, N_2 q_\chi\,\,,
\ee
where $N_2$ gives the number of fermions of opposite charge it needs to accrete to become neutral. We now discuss in turn estimates of the discharge time for a dark charged black hole, and then how it could acquire dark charge in the first place. These issues all hinge on the possibility that a charge $q_\chi$ with mass $m_\chi$ is able to be absorbed by a black hole with charge $Q$ and mass $M$. A simple (classical) comparison of the associated gravitational attraction and possible dark repulsion yields the condition 
\be
q_\chi Q \, \sim \, m_\chi M \,\,\,\,\,\, \Rightarrow \,\,\,\,\,\, \frac{Q}{M} \, \sim \, \frac{m_\chi}{q_\chi}\,\,.
\ee
In the Standard Model, this condition is badly violated unless $Q \sim 0$. However, for a dark sector with
\be \label{darkchargechoice}
m_\chi \, \sim \, m_p, \,\,\,\,\,\,\,{\rm and} \,\,\,\,\,\,\, q_\chi \, \sim \, 10^{-18} \, e \,\,\,
\ee
where $e$ is the charge of the electron and  $m_p$ is the mass of a proton, one can obtain $Q \sim M$. 

Assuming that the SMBH is in an environment with density $\rho$ of oppositely charged dark fermions moving with velocity $v$, and that the accretion follows the Bondi rate for collisionless fluid, \cite{Cardoso:2016olt} estimates the discharge time to be
\be
\tau_{\rm discharge} \, \sim \, \left(\frac{v}{220 \,{\rm km/s}}\right)^3\left(\frac{0.4 \,{\rm GeV/cm}^3}{\rho}\right)\left(\frac{10 M_\sun}{M_2}\right)\left(\frac{m_\chi}{m_p}\right)\left(\frac{e}{q_\chi}\right) \,\, {\rm yr}\,\,.
\ee
With the choice of dark sector parameters given in Eq.~\eqref{darkchargechoice}, this results in a large discharge time. Another choice of the mass and charge could be $m_\chi \,\sim \, 10^{-3} m_p$ and $q_\chi \sim 10^{-21}\,e$, which leaves the discharge time unchanged. If the black hole is located in an environment where $\rho \sim 0$, the discharge time can be even longer. We note that the above estimate for the discharge time assumes an accretion rate given by
\be
\frac{dM_2}{dt} \, \sim \, M^2_2 Q_{\rho}\,\,,
\ee
where $Q_\rho \sim \rho/v^3$ is the dark fermion phase space density \cite{Peirani:2008bu}. Moreover, it assumes that the entire density $\rho$ is constituted by oppositely charged dark fermions, which is the most conservative assumption. 

The question of how a SMBH can acquire dark charge in the first place is a difficult one and is tied to the formation mechanism of SMBHs, which is unknown. It may be easier to answer this question  for the case of stellar black holes first. Classic calculations \cite{1926ics..book.....E} indicate that the charge of a typical star at birth may be given by
\be \label{bhcollap}
\left(\frac{Q}{M}\right)_{\rm birth} \, \sim \, \frac{m_\chi}{q_\chi}\,\,
\ee
under the assumption that the black hole is formed by the collapse of dark charged particles. For the Standard Model, the charge obtained from Eq.~\eqref{bhcollap} is miniscule, and is promptly discharged as well. For dark sectors with $\frac{m_\chi}{q_\chi} \sim 1$, however, one can obtain $\left(\frac{Q}{M}\right)_{\rm birth} \sim 1$, and the discharge time can also be large.

For a SMBH, the situation is less clear. The proliferation of SMBH observations with $z > 5$ raises the question about their origin, which  is a mystery (we refer to \cite{Inayoshi:2019fun} for a review). One possibility is the existence of intermediate mass black hole seeds with $M_2 \sim 10^{3-5} M_\sun$ at $z \sim 20$, along with a population of stellar black holes. If these SMBH seeds are formed from the collapse of a dark matter halo, calculations similar to the one sketched above may apply to their initial charge. The subsequent accretion history of the seed is expected to follow mean Eddington rates with episodes of super-Eddington accretion \cite{1988ApJ...332..646A, Volonteri:2014lja, Inayoshi:2015pox, Hardcastle:2020vhm}:
\be
M_2(t) \, \sim \, M_{\rm seed} \times e^{t/\tau_{\rm Edd}}\,\,,
\ee
where $\tau_{\rm Edd} = 45 $ Myr is the Eddington timescale. Accretion under these circumstances can grow $\sim \mathcal{O}(10-100) M_\sun$ seeds into black holes with mass $\sim 10^3 M_\sun$ by $z=18$ and $\sim 10^8 M_\sun$ by $z=7$. 

Assuming that the growth of the black hole mass between $z \sim 30$ to $z \sim 7$  proceeds by the accretion of equal numbers of positively and negatively charged dark particles,  the final charge of the SMBH would be equal to the initial charge, while the mass is enhanced by orders of magnitude. This would yield, for example, $(Q/M)_{\rm final} \sim 10^{-5}$, assuming that $(Q/M)_{\rm birth} \sim 1$. On the other hand, preferential accretion of one charge over the other could produce $(Q/M)_{\rm final} \sim 1$.

\bibliographystyle{JHEP}
\bibliography{reference}

\end{document}